\documentstyle[twocolumn,aps,epsfig,floats]{revtex}

\begin{document}

\draft \tolerance = 10000

\setcounter{topnumber}{1}
\renewcommand{\topfraction}{0.9}
\renewcommand{\textfraction}{0.1}
\renewcommand{\floatpagefraction}{0.9}

\twocolumn[\hsize\textwidth\columnwidth\hsize\csname
@twocolumnfalse\endcsname

\title{About the Dependence of the Currency Exchange Rate at Time
and National Dividend, Investments Size, Difference Between Total Demand
and Supply}
\author{L.Ya. Kobelev, O.L. Kobeleva, Ya.L. Kobelev \\
Department of Problems of Extreme Influence on Matter, \\ Department  of
Low Temperatures Physics, Ural State University,\\ Lenina av., 51,
Ekaterinburg, 620083, RUSSIA}

 \maketitle

\begin{abstract}
The time dependence of the currency exchange rate $K$  treated as a
function of national dividend, investments and difference between total
demand for a goods and supply is considered. To do this a proposed earlier
general algorithm of economic processes describing on the basis of the
equations for $K$ like the equations of statistical physics of open
systems is used. A number of differential equations (including nonlinear
ones too) determining the time dependence of the exchange rate (including
oscillations) is obtained.
\end{abstract}
\vspace{1cm}

]

1. L.Ya.Kobelev et al. \cite{Kob1} in order to describe economic phenomena
offered a system of  nonlinear differential equations based on the
mathematical methods of statistical physics of open systems
(\cite{Klim}).(see also \cite{Byst1}-\cite{Meod}). The present paper is
devoted to consideration of a model example of calculation of time
dependence of national currency  exchange rate treated as a function of
national dividend $F$, investments C and total difference between demand
and supply (with all the values being measured in national currency
units). The modality of the example considered is in neglecting of
dependencies of national currency upon the others, not mentioned economic
variables, so the main purpose of this paper is not so to elucidate real
economic laws and regularities (although this also takes place) as to
illustrate the possibilities of the offered in the named paper method if
applied to a concrete problem.

2. Thus, to show the advantages of the method used consider the dependence
of the currency exchange rate upon investments $C$, difference between
total demand and supply P, and national dividend $F$. Let K be a variation
of the currency exchange rate per time unit. In this case the equations
describing the exchange rate as a function of $C$, $P$ and $F$ takes the
form of the simplified equations (3), (4), (7) of the paper \cite{Kob1}
\begin{equation}\label{1}
\frac{dK}{dt}=\frac{\partial K}{\partial t}+\frac{\partial K}{\partial
C}\frac{dC}{dt}+ \frac{\partial K}{\partial P}\frac{dP}{dt}+
\frac{\partial K}{\partial F}\frac{dF}{dt}=I_C
\end{equation}
\begin{eqnarray}\label{2}
\nonumber &&I_C=\varphi(K)+\frac{\partial}{\partial
C}(D_C\frac{\partial}{\partial C}K)+\\ && +\frac{\partial}{\partial
p}(D_P\frac{\partial}{\partial P}K)+\frac{\partial}{\partial
F}(D_F\frac{\partial}{\partial F}K)
\end{eqnarray}
The equations describing time dependencies of $C$, $P$, $F$ and the
dependence of the currency exchange rate upon them write down as follows:
\begin{equation}\label{3}
\frac{dK}{dx_i}=F_i(x_i,K)+\frac{\partial}{\partial x_j}[D_{j\beta}
\frac{\partial}{\partial x_\beta}(x_i-A_jx_i)]
\end{equation}
\begin{equation}\label{4}
\frac{dx_i}{dt}=\varphi(x_j,\dot{x_j},K,\dot{K}, \frac{\partial K}
{\partial x_j})
\end{equation}
where $i=1,2,3$ ($x_1=C$, $x_2=P$, $x_3=F$),  $F_i$  and $\phi$ are
nonlinear functions of their own arguments. In eqs. (\ref{2})-(\ref{3})
the terms containing $D_C$, $D_P$ and $D_F$ coefficients describe
"diffusion" in the space of the exchange rate alteration variables $C$,
$P$, $F$, i.e., the distribution of $K$ over these variables, and the
$\phi(K)$ terms describe the dependence of $\frac{dK}{dt}$ upon the
processes regulating the velocity of the exchange rate alteration (in
particular, a currency exchange rate relaxation, bistable states existence
etc.).

3. In the simplest case chose $_C=0$, that is, consider the equation
\begin{equation}\label{5}
\frac{dK}{dt}=\frac{\partial K}{\partial t}+\frac{\partial K}{\partial
C}\frac{dC}{dt}+ \frac{\partial K}{\partial P}\frac{dP}{dt}+
\frac{\partial K}{\partial F}\frac{dF}{dt}=0
\end{equation}
The equality $I_C$=0 corresponds to neglecting by the currency exchange
rate alteration smoothing processes influence on $\frac{dK}{\partial t}$.
The $\frac{\partial K}{\partial C}$, $\frac{\partial K}{\partial P}$ and
$\frac{\partial K}{\partial F}$ derivatives in (\ref{5}) are determined,
in difference with kinetic theory in statistical physics, by eqs.(\ref{3})
and the derivatives $\dot{x}_i$ are governed by (\ref{4}).

4. Consider now, for the sake of simplicity, the case when the influence
of $P$ and $F$ on the currency exchange rate alteration is very small and
may not be taken into account. Then
\begin{equation}\label{6}
\frac{\partial K}{\partial t}+\frac{\partial K}{\partial C}\frac{dC}{dt}=0
\end{equation}
To define $\frac{\partial K}{\partial C}\frac{dC}{dt}$ we use a reduced
equations (\ref{1})-(\ref{4})
\begin{equation}\label{7}
\frac{dC}{dt}=a\int_0^t{K(t)dt}
\end{equation}
\begin{equation}\label{8}
\frac{\partial K}{\partial C}=\alpha
\end{equation}
where $\alpha$ in (\ref{7}) is a velocity of changing in investments
attributed to the sum of the currency exchange rate alteration over the
time interval of $t$ and $\alpha$ in (\ref{8}) is the changing in the
currency exchange rate alteration per investments unit. \\ Assume then
that these values do not vary significantly, that is $\alpha=const$ and
$\frac{dC}{dt}= b=const$. In this case we will have $\frac{\partial
K}{\partial t}=K_1=ab$ and $K=K_1t+K_2$ ($K_2=const.$). So $K$ will
increase or decrease, depending upon the sign of $K_1$. In case of
(\ref{7}), differentiating (\ref{6}) in time and using (13) from paper
\cite{Kob1} and (\ref{8}) yield
\begin{eqnarray}\label{9} \nonumber
\frac{\partial^2 K}{\partial t^2}+\omega_0^2K=0
\end{eqnarray}
where $\omega_0=\sqrt{a\alpha}$. The solution of (\ref{9}) has the form
\begin{eqnarray}\label{10} \nonumber
K=K_0\sin\omega_0t
\end{eqnarray}
where $K_0$ is constant. So with the assumption made, the currency
exchange rate alteration $K$ varies periodically in time with the
frequency of $\sqrt{a\alpha}$. If the dependence of $K$ upon the
investments made is weak, the frequency of oscillations will  be small and
K will take the form ($\omega_0 t \ll 1$)
$$K=K_0\sqrt{a\alpha}t+...\qquad\qquad \sqrt{a\alpha}t \ll 1$$ At times
large enough, nevertheless, the periodicity of the currency exchange rate
varying with time will be observed.

5. Introduce now into the right-hand side of eq.(\ref{5}) a parameter
characterizing the deviation of $K$ from an equilibrium:
$I_c=\frac{K-K_0}{\tau(K)}$, where $\tau(K)$ is the relaxation time for
the currency exchange rate alteration if diverged from a stationary state
$K_0$ (in general case $\tau$ is a function of $K$, $\dot{K}$, $C$, $F$
etc.)\\Then instead of (\ref{6}) we have
\begin{equation}\label{11}
\frac{\partial K}{\partial t}+\frac{\partial K}{\partial C}\frac{dC}{dt}=
\frac{K-K_0}{\tau(K)}
\end{equation}
Assuming $K_0=const$ and $\tau=const$ differentiating in time, if eqs.
(\ref{7}) and (\ref{8}) are valid, yields:
\begin{eqnarray}\label{12} \nonumber
\ddot{K}+\omega_0^2K-\frac{1}{\tau}\dot{K}=0
\end{eqnarray}
whose solution is
\begin{eqnarray}\label{13} \nonumber
K=k_0 e^{-\frac{t}{\tau}} \cos{\omega_0t}
\end{eqnarray}
In this case the oscillations of the currency exchange rate changing will
damp tending to zero when $t \ll \tau$.

6. Take into account in equation (\ref{11}) the difference $P$ between
total demand and supply influence on the exchange rate alteration,
determining $\frac{dP}{dt}$ and $\frac{\partial K}{\partial P}$ from the
equations (a special case of (\ref{1}) and (\ref{4}))
\begin{eqnarray}\label{14} \nonumber
a)\;\frac{dP}{dt}=\gamma\dot{K},\quad \frac{\partial K}{\partial P}=
\beta K^2
\end{eqnarray}
\begin{eqnarray}\label{15} \nonumber
b)\;\frac{dP}{dt}=\beta K^2,\quad \frac{\partial K}{\partial P}=
\gamma\dot{K}
\end{eqnarray}
with linear (a) or nonlinear (b) dependence of $\frac{dP}{dt}$ upon
$\dot{K}$ or $K$ correspondingly. Then instead of (\ref{11}) we get
\begin{equation}\label{16}
\frac{\partial K}{\partial t}+a\alpha\int_0^t{K(t')dt'}+ \beta\gamma
K^2\dot{K}=\frac{K-K_0}{\tau(K)}
\end{equation}
or, after differentiating with respect to time
\begin{equation}\label{17}
(1+\beta\gamma K^2)\ddot{K}-(\frac{1}{\tau}-2\beta\gamma K\dot{K})\dot{K}+
\omega_0^2K=0
\end{equation}
If $\beta \gamma K^2 \ll 1$ and $2\beta \gamma K \dot{K} \gg
\frac{1}{\tau}$, then (\ref{16}) reduces to
\begin{equation}\label{18}
\ddot{K}+2\delta(K\dot{K})\dot{K}+\omega_0^2K=0
\end{equation}
where
\begin{eqnarray}\label{19} \nonumber
\delta(K\dot{K})=\beta\gamma K\dot{K}
\end{eqnarray}
Eq. (\ref{17}) is an equation of oscillations with a nonlinear friction
$2\delta(K\dot{K})$ and its solutions contains all the peculiarities of
nonlinear systems. Take into consideration in (\ref{16}) the term
$\frac{\partial}{\partial P}(D_P \frac{\partial}{\partial P})K$, assuming
$D_P=const$. This yields a change in coefficient at in (\ref{18}) in the
case (a)
\begin{eqnarray}\label{20}
(1+\beta\gamma K^2)\ddot{K}&-&(\frac{1}{\tau}-2\beta\gamma K\dot{K}+
D_P\beta^2K^2)\dot{K}+\\ \nonumber &+&\omega_0^2K=0
\end{eqnarray}
and appearance of a term containing $\frac{d^3K}{dt^3}$ in the case (b)
\begin{eqnarray}\label{21} \nonumber
&& \gamma^2D_P^2\frac{d^3K}{dt^3}-(1+\beta\gamma K^2)\ddot{K}- \\
\nonumber  &&-2\beta(K\dot{K})\dot{K}-\omega_0^2K=0
\end{eqnarray}

7. One may take into account in (\ref{16}) the terms containing
$\frac{\partial K}{\partial F}\frac{dF}{dt}$ assuming, for example,
$D_F=const$, and (a special case of (\ref{1})-(\ref{4}))
\begin{eqnarray}\label{22}
&& a)\;\frac{\partial K}{\partial F}=b\int_0^t{K(t')dt'},\quad
\frac{dF}{dt}=d=const \\ \nonumber && b)\;\frac{\partial K}{\partial
F}=d=const,\quad \frac{dF}{dt}= \int_0^t{K(t')dt'}
\end{eqnarray}
Eqs. (\ref{20}) then takes the from ((\ref{22}a) case)
\begin{eqnarray}\nonumber
&&(1+\beta\gamma K^2)\ddot{K}-(\frac{1}{\tau}-2\beta\gamma K\dot{K}+
D_P\beta^2K^2)\dot{K}+ \\ &&
[(\omega_0^2+bd)K-D_F\beta^2\int_0^t{K(t')dt'}]=0
\end{eqnarray}
and after differentiating in $t$
\begin{eqnarray}\label{23} \nonumber
&&\nonumber (1+\beta\gamma
K^2)\frac{d^3K}{dt^3}-(\frac{1}{\tau}-2\beta\gamma K\dot{K}
+D_P\beta^2K^2)'\dot{K}+\\ &&+(\frac{1}{\tau}-2\beta\gamma K\dot{K}
+D_P\beta^2K^2)\ddot{K}+ \\ \nonumber &&
+[(\omega_0^2+bd)\dot{K}-D_F\beta^2K(t)]+ 2\beta\gamma K\dot{K}\ddot{K}=0
\end{eqnarray}
Using (\ref{22}b) one obtains
\begin{eqnarray}\label{24} \nonumber
&& (1+\beta\gamma K^2)\ddot{K}-(\frac{1}{\tau}2\beta\gamma K\dot{K}+
D_P\beta^2K^2)\dot{K}+ \\ \nonumber && +(\omega_0^2+bd)K=0
\end{eqnarray}
Consider finally a case of nonlinear dependence of $I_C$ upon $K$. Let,
remaining the assumptions of the previous paragraphs concerning the
dependencies of $K$ upon $D$, $C$ and $F$, $I_C$ has the form
\begin{eqnarray}\label{25}
I_C=\int\limits_0^t{}(l_0&-&lK^2)Kdt+\frac{K-K_0}{\tau}+
D_P\frac{\partial^2K}{\partial P^2}+ \\ \nonumber &&
+D_C\frac{\partial^2K}{\partial C^2}+D_F\frac{\partial^2 K}{\partial F^2}
\end{eqnarray}
The choice of (\ref{25}) corresponds to nonlinear type of $K(t)$
relaxation. Substituting $I_C$ from (\ref{25}) (e.g., for (\ref{22}b)
case) into (\ref{2}) yields (using the proper equations for derivatives of
$K$, $C$, $P$, $F$):
\begin{eqnarray}\label{26}
(1+\beta\gamma K^2)\ddot{K} && +(2\beta\gamma K\dot{K}-D_P\beta^2K^2-
\frac{1}{\tau})\dot{K}+ \\ \nonumber && +[(\omega_0^2+bd)-(l_0-lK^2)]K=0
\end{eqnarray}
Eq. (\ref{26}) has a bifurcation point at
\begin{eqnarray}\label{27} \nonumber
K=\pm\frac{1}{\sqrt{l}}\sqrt{l_0-\omega_0^2-bd}
\end{eqnarray}
(which corresponds to the appearance of a bistable state for the currency
exchange rate alteration) and a number of other interesting peculiarities
as well (in particular, at $K=\pm\sqrt{\frac{1}{|\gamma\beta|}}$,
$(\gamma\beta<0)$,
\begin{eqnarray}   \nonumber
&& at K=\frac{1}{D_P\beta^2}(-\beta\gamma\dot{K}\pm
\sqrt{(\beta\gamma\dot{K})^2+\frac{2D_P\beta^2}{\tau}},\\ &&\nonumber at
\dot{K}=\frac{D_P\beta^2K\pm\sqrt{D_P\beta^2K^2-8\beta\gamma
(\omega_0^2+bd-l_0-lK^2)}}{4\gamma\beta},
\end{eqnarray}
etc.)

8. Note, that wide opportunities to chose the equations for
$\frac{\partial K}{\partial C}$, $\frac{\partial K}{\partial P}$,
$\frac{\partial K}{\partial F}$ and $\frac{dC}{dt}$, $\frac{dP}{dt}$,
$\frac{dF}{dt}$ and $I_C$ are not exhausted by the selection used in the
example considered and the concrete forms of the equations are determined
by the state of  the economic processes and correlations between them. For
example, one may treat as $x_i$ variables in the case considered  realized
$Y_e$ and produced $Y_l$ gross national products, the total sum of money
in use, the number of workable population $N$, absolute level of
unemployment $\delta N$, the part of gross national product used by the
state (these variables were used by Bystrai (\cite{Byst1}) while defining
the economic entropy.
 \\ \\ \\ \\ \\ \\
\begin{center}
{\bf CONCLUSION}
\end{center}

A general algorithm of describing economic processes basing on the
equations of statistical physics of open systems developed by Kobelev
L.Ya. et al. \cite{Kob1} was used to describe the time dependence of the
national currency exchange rate as a function of national dividend,
investments size and difference between total demand for the goods and its
supply. A number of nonlinear differential equations describing the time
dependence of exchange rate (in particular, oscillations in different
cases) were obtained.

\end{document}